# High chemical activity of a perovskite surface: Reaction of CO with $Sr_3Ru_2O_7$

Bernhard Stöger[1], Marcel Hieckel[1], Florian Mittendorfer[1], Zhiming Wang[1], David Fobes[2], Jin Peng[2], Zhiqiang Mao[2], Michael Schmid[1], Josef Redinger[1], Ulrike Diebold[1]

[1] Institute of Applied Physics and Center for Computational Materials Science, Vienna University of Technology, Vienna, Austria

[2] Department of Physics and Engineering Physics, Tulane University, New Orleans, LA, USA

PRL Abstract: 600 characters

Adsorption of CO at the $Sr_3Ru_2O_7$(001) surface was studied with low-temperature STM and DFT. *In-situ* cleaved single crystals terminate in an almost perfect SrO surface. At 78 K CO first populates impurities and then adsorbs above apical surface O with a binding energy $E_{ads}$ = -0.7 eV. Above 100 K this physisorbed CO replaces the surface O, forming a bent $CO_2$ with the C end bound to the Ru underneath. The resulting metal carboxylate (Ru-COO) can be desorbed by STM manipulation. A low activation (0.2 eV) and high binding (-2.2 eV) energy confirm a strong reaction between CO and regular surface sites of $Sr_3Ru_2O_7$; likely this reaction causes the 'UHV ageing effect' reported for this and other perovskite oxides.

Complex ternary perovskite oxides are increasingly used in solid oxide fuel cells and catalysis [1-5] and in emerging devices based on superconductivity, ferroelectricity, magnetoresistance, and other properties that can be tuned by external parameters such as doping, fields, pressure, and temperature [6]. In view of these applications, it is highly desirable to obtain a better understanding of their surface chemical properties at the atomic level. Recent scanning tunneling microscopy (STM) investigations of molecular adsorption provide valuable fundamental insights [7-9], yet compared to binary oxide surfaces [10, 11] the level of understanding is still woefully inadequate.

The lack of fundamental studies is largely due to the difficulty of preparing perovskite surfaces with a well-defined structure. Preferential sputtering, surface segregation and polarity effects often cause reconstructions [12]. Almost ideal, bulk-terminated surfaces are achieved by in-situ cleaving of layered perovskites; such systems are often used in angle-resolved photoemission spectroscopy (ARPES) and STM measurements of the electronic structure [13-15]. Even for these model systems it is difficult to identify the image contrast in STM and the defects present at the surface [16, 17], however. The present, combined experimental and theoretical study reports the controlled adsorption of CO, the principal molecule for probing surface chemistry of oxides in spectroscopic studies [18].

As a sample $Sr_3Ru_2O_7$ was chosen. This is the ($n$ = 2) member of the n-layered ruthenate Ruddlesden-Popper series $Sr_{n+1}Ru_nO_{3n+1}$, which consists of $n$ perovskite-like $SrRuO_3$ layers separated by two layers of SrO [see Fig. 1(a)]. Materials of this class cleave easily along the (001) plane between the adjacent SrO layers without breaking the $RuO_6$ octahedra. Cleaving $Sr_3Ru_2O_7$ in ultrahigh vacuum (UHV) creates a non-polar, non-reconstructed SrO-terminated surface, and thus a well-defined starting point for surface experiments. There are good reasons to investigate the surface chemistry of strontium ruthenates. Due to its high electrical conductivity and interesting physical properties, $SrRuO_3$ is a preferred electrode material in oxide-based electronic devices [19]; reaction with hydrocarbons in the ambient air render the material thermally unstable [20, 21]. Owing to the volatility of higher ruthenium oxides, $SrRuO_3$ films grow with a SrO termination [22], results from cleaved $Sr_3Ru_2O_7$ samples should thus be representative for the surfaces of $SrRuO_3$ and other strontium-based perovskite films. In addition, somewhat puzzling effects have been reported for $Sr_2RuO_4$ and $Sr_3Ru_2O_7$ samples cleaved in ultrahigh vacuum (UHV), e.g., surface properties that are dependent on cleaving temperature [17], as well as an 'ageing' of the surface when the temperature of the sample is cycled [13] or when it is simply kept in vacuum at low temperatures [23]. Carbon monoxide is one of the main constituents of the residual gas in UHV. Here we show that the surface degrades by formation of carbon based species, which can be desorbed by inelastic electronic transitions.

The present work shows that CO interacts strongly with the SrO surface of $Sr_3Ru_2O_7$. The adsorbed molecule reacts with a surface O and forms a COO entity that binds strongly with the underlying Ru. This species can be removed with the STM tip, suggesting that desorption via electronic excitations should be good way to clean the surface.

The experiments were carried out in a two-chamber UHV system with base pressures of $2\times10^{-11}$ and $6\times10^{-12}$ mbar in the preparation chamber and the STM chamber, respectively. A low-temperature STM (commercial Omicron LT-STM) was operated at 78 K in constant-current mode using an electro-chemically etched W-tip, with the STM bias voltage applied to the sample. High-quality $Sr_3Ru_2O_7$ single crystals were grown in a two-mirror image floating zone furnace, for details see ref. [24]. The samples were fixed on Omicron sample plates with conducting silver epoxy (EPO-TEK H21D), and a metal stud was glued on top with another epoxy adhesive (EPO-TEK H77). The crystals were cleaved by removing the metal stud with a wobble stick. Cleaving was performed in the analysis chamber at temperatures between 100 K and 300 K; in agreement with ref. [25] the cleaving temperature did not influence the results. After cleaving the sample was immediately transferred into the cold STM; the first images were usually obtained within 30 minutes after cleaving.

The spin-polarized density functional theory (DFT) calculations were performed with the Vienna Ab-initio Simulations Package (VASP) in the PAW framework [26], using the Perdew Burke Ernzerhof (PBE) exchange-correlation functional [27]. Test calculations with an enhanced onsite Coulomb interaction (DFT+U) with an U-J value of up to 4 eV [28] yield similar results for the adsorption energies. The surface was modeled by a single ferromagnetic (4×4) $Sr_3Ru_2O_7$ double layer slab terminating at the cleavage plane [Fig. 1]. Convergence tests for the bare $Sr_3Ru_2O_7$ surface show a change in the surface energy of ~0.01 eV/Å$^2$ going from a single to a double bilayer. The uppermost three layers were fully relaxed. Brillouin zone integration was performed on a 2×2×1 Monkhorst Pack k-point mesh. The reaction barriers were identified by the dimer method [29] with a subsequent, explicit verification of the reaction pathways. The bonding analysis (COOP, [30]) is based on the local orbital phase factors using the PAW projectors.

The surfaces of cleaved $Sr_3Ru_2O_7$ are flat with terraces up to a few μm², separated by 1.1 nm-high steps. They typically contain <0.5 percent of a monolayer (ML) of defects, see Fig. 2. These defects are attributed to bulk impurities rather than artifacts of the cleaving process or adsorbates from the residual gas [31]. In $Sr_3Ru_2O_7$ the $RuO_6$ octahedra are rotated clockwise and anticlockwise by 8.1° at 90 K [32], see the top view in Fig. 1(b).

The first principles calculations confirm that the Sr and O atoms of the SrO layer are imaged as bright protrusions and dark depressions in STM, respectively, in agreement with previous work [33]. The rotation of the octahedral units yields two inequivalent Sr and apical O atoms at the surface [Fig. 1(b)]. At 78 K and the tunneling conditions applied here the inequivalent octahedra of the unit cell cannot be distinguished most of the time (depends strongly on the tip quality). Defects appear different depending on their lattice site [15, 16, 34].

Carbon monoxide was dosed in steps of 0.0015 L (1 L = $10^6$ torr s) while the sample was in the STM at 78 K [see Fig. 2]. Imaging the surface before and after exposure allows the identification of the adsorption site. The CO molecule adsorbs first at the surface defects, indicating transient mobility of the adsorbate. Once all defect sites are saturated, CO adsorbs at the bare SrO surface.

In STM the as-dosed CO appears as a bright spot on the clean surface [red arrow in Fig. 3(a)], centered on top of an oxygen atom of the SrO layer. This configuration is consistent with a DFT-derived geometry [Fig. 3(c)], where the C bonds downwards to a surface oxygen atom, and the O end points towards a Sr bridge site. Upon adsorption the C-O bond length increases, from a calculated value of 1.14 Å in the gas phase to 1.27 Å in the adsorbed state. The distance between the carbon atom and the surface oxygen atom is 1.35 Å. The adsorbed CO molecule leads to a local distortion of the lattice; as the bond length between the surface O and the Ru atom below increases by 0.2 Å. An STM simulation based on this adsorption geometry agrees well with the experimental result; see Fig. 3(c).

In addition to these relatively weakly bound CO molecules the STM image in Fig. 3(a) shows three dark crosses, each with one thicker and one thinner arm. With DFT these crosses are identified as a chemisorbed configuration ($E_{ads}$ = -2.17 eV). The carbon atom of the CO molecule is incorporated into the surface layer by replacing the apical oxygen atom and forming an adsorbed Ru-COO species, best classified as a metal carboxylate [Fig. 3(d)]. The Ru atom is pulled upwards by 0.33 Å to allow for a C-Ru bond length of 2.03 Å. The two oxygen atoms of the COO-group point towards two Sr-bridge positions with an O-C-O angle of 118.7° [see Fig 3(d)], significantly smaller than the angle of 133° predicted for a charged $CO_2$ molecule [35]. The COO molecule is symmetric; both C-O bonds are elongated to a value of 1.3 Å due to the partial occupation of antibonding intramolecular states, which is enhanced by the bending of the molecule (see also Fig. 4, below). The formation of the carboxylate also causes a local distortion of the surrounding lattice, in particular a tilt of the neighboring $RuO_6$ octahedra that is more pronounced for octahedra aligned with the COO axis [~6°, see Fig 3(e)]. The calculated and experimental STM images again agree well.

The transformation of the physisorbed CO to the Ru-COO carboxylate was simulated with DFT. The reaction proceeds in a concerted mechanism, where the O-C-O complex rotates by 90°, and simultaneously breaks and forms a O-Ru and C-Ru bond, respectively [Fig. 3]. This process has a surprisingly small energy barrier of only 0.17 eV, see the potential energy diagram in Fig. 3(f). This activation barrier is overcome by either annealing the sample to 100 K, or by scanning the STM tip at a bias of ±1 V across the physisorbed configuration. This is shown in sequence Figs. 3(a, b) where a physisorbed CO, marked in red, transforms into a CO cross. The blue arrows in Fig. 3(b) mark two CO crosses that are rotated by 90°. These two features correspond to the two symmetrically equivalent adsorption configurations, where the carboxylate O-C-O-axes are rotated by 90°. This rotation, which has a calculated $E_{barr}$ = 0.44 eV [Fig. 3(e)], can also be induced by scanning at +2.4 V (not shown).

The strong interaction between CO and the SrO layer of $Sr_3Ru_2O_7$ is also reflected in a high initial sticking coefficient. When 0.02 L CO was dosed, each CO molecule that hit the surface adsorbed. As mentioned above, the CO molecules are initially mobile on the surface, which allows for intermolecular interactions. Exposure of CO on UHV-cleaved $Sr_2RuO_4$ samples results in the same O-C-O 'crosses', albeit the physisorbed precursor was not observed. This n = 1 member of the Ruddlesden-Popper series is even more reactive; exposure to 25 L CO at RT resulted in a coverage of 3.0% and 8.5% ML for $Sr_3Ru_2O_7$ and $Sr_2RuO_4$, respectively.

When a CO-covered $Sr_3Ru_2O_7$ sample was annealed to 420 K no desorption was observed. This is not surprising considering a calculated binding energy of -2.2 eV. In this study sample heating was limited to avoid outgassing of the glue, thus it was not tested experimentally in what form the CO would leave the surface. DFT suggests that the molecule should desorb as CO rather than as $CO_2$. In PBE calculations the oxidation of CO to $CO_2$ results in an energy gain of 3.3 eV, but desorption creates a surface oxygen vacancy, which costs 3.8 eV.

The CO can be cleaned off locally with the STM tip. Applying a bias voltage of ± 0.4 V removes the physisorbed precursor [see Supplement], and a bias voltage of +2.7 V causes the chemisorbed CO 'crosses' to disappear from the scanned area [Fig. 4(a)]. Scanning at negative sample bias voltages does not remove the chemisorbed CO molecules.

Possibly the tip-induced removal of the physisorbed, weakly-bound precursor happens via excitation of stretching vibrations through inelastic tunneling. The DFT calculations indicate a molecule-surface vibrational mode at ~120 meV, consistent with the observation that the precursor is removed at bias voltages between ±(0.2 V – 0.4 V). On the other hand, the chemisorbed, OCO-like species is most likely removed by electron capture into antibonding orbitals. DFT predicts that the lowest antibonding O-CO molecular orbital of the carboxylate, the $2b_1$ orbital, is centered around +2.4 eV [see Fig. 4(b)]. A detailed analysis of the orbital (wavefunction) phase factors of the respective atoms as shown in Figs. 4(c, d), reveals that this state is antibonding with respect to both, the substrate (lower panel) and the OCO molecule. Populating this orbital will facilitate desorption, as well as the dissociation of the molecule by weakening the C-O bond. This could explain the observed CO removal in STM. Experiments at higher bias voltages suggest that field-induced processes may also play a role.

The experimental and computational results clearly point towards a strong interaction between CO and SrO-terminated perovskite ruthenates that needs to be taken into account even for experiments under the most pristine conditions. How general are these results? Is this high reactivity a characteristic property of the terminating SrO layer? What is expected for other terminations, in particular for Ca or Ba, which are often used as the A cation in perovskites? And what is the role of the B cation; is a Ru-based perovskite particularly reactive, or is the observed reaction with lattice O to be expected for other perovskites as well?

The interaction between CO and binary oxides of earth alkali metals was reviewed in ref. [18]. The reactivity increases dramatically with the basicity of the oxide, i.e., MgO < CaO < SrO. Interestingly, a reaction between lattice O and CO was postulated as the first step of a CO polymerization process; it is well possible that this critical, initial species is the O-C-O entity identified in this work. For MgO, CO reacts only with highly undercoordinated O sites, while for SrO it was conjectured that regular sites at facet planes should be involved [36]. This agrees with the observation that CO readily adsorbs on perfect $Sr_3Ru_2O_7$ and $Sr_2RuO_4$ surfaces. Under UHV conditions CaO is relatively inert unless activated by extrinsic dopants [37]. Preliminary experiments on $Ca_3Ru_2O_7$ however point towards strong CO adsorption, although it needs to be investigated in more detail if the reactivity of Sr-terminated and Ca-terminated ruthenates is indeed comparable, and if the same species form.

Concerning the influence of the B-site, it is instructive to compare the reactivity of binary transition metal oxides to CO. Generally, a high activity is a direct consequence of a high reducibility (i.e. a low formation energy of oxygen vacancies), as this facilitates the removal of O from the surface. On $RuO_2$ (110), CO adsorbs at the undercoordinated Ru atom, reacts with the neighboring bridging oxygen atom, and desorbs as $CO_2$ [38]. $TiO_2$(110), which has the same rutile structure, is far less reactive, and CO adsorbs weakly below 200 K [39]. We therefore predict a higher reactivity for Sr-based perovskites with e.g., B=Mo, Ir, Mn than for more electropositive cations (B=Ti). A weak interaction was indeed reported for $CO/SrTiO_3$ [40] but, as pointed out above, the surface termination of such sputtered/annealed perovskite samples is unfortunately often not well-defined.

In conclusion, the high reactivity found in the present study needs to be considered when studying the properties of perovskite surfaces even under stringent UHV conditions. It has been recognized early on that a 'degradation' of high-Tc superconductors occurs in UHV [41, 42] due to gas adsorption. Our results show that the interaction with CO clearly plays a major role for strontium ruthenate surfaces. An analysis of the CO adsorption configuration shows a pronounced change in the local structure, which is intimately connected with electronic and magnetic properties in strongly correlated materials [33]. Indeed it has been proposed that 'aging' a sample can be utilized to suppress 'surface states' [13] in ARPES measurements. We also suggest how to restore a high-quality sample. While heating an adsorbate-covered $SrRuO_3$ can result in decomposition of the sample [20, 21], our STM and DFT results suggest that a gentle removal of CO should be possible via electronic excitations. Finally, these investigations of an (almost) perfect SrO-terminated surface have directly identified an oxidized CO species that has been postulated to play a major role in the surface chemistry of earth alkali oxides [18].

This work was supported by the Austrian Science Fund (FWF, Project F45) and the ERC Advanced Grant 'OxideSurfaces'. The work at Tulane is supported by the NSF under grant DMR-1205469. The crystals structures were plotted using VESTA [43]. VisIt [44] was used to visualize the simulated STM data.

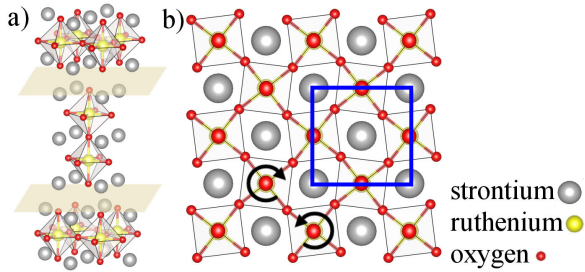

**Fig 1.** (a) Crystal structure of $Sr_3Ru_2O_7$. The cleavage planes (pale brown) between two SrO layers mark the weakest bonds. (b) Surface structure. The top surface layer contains apical O (bigger red dots) and Sr atoms (grey). The $RuO_6$ octahedra are rotated alternatingly clockwise and counterclockwise. The bluesquare marks the orthorhombic unit cell.

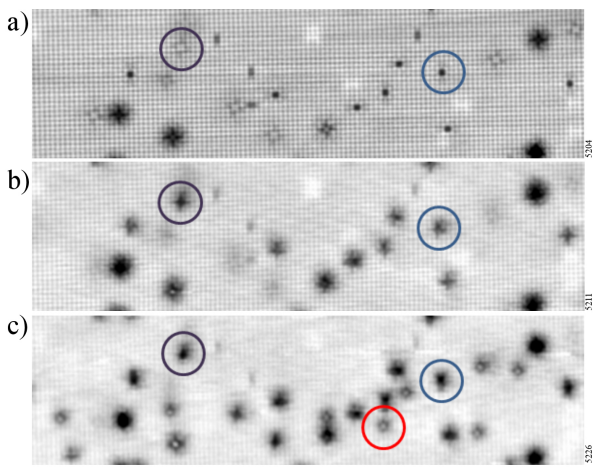

**Fig 2. Adsorption of CO on $Sr_3Ru_2O_7$.** STM topographies: 45×12 nm², $T$ = 78 K (a) $U_{sample}$ = +0.05 V / $I_t$ = 0.15 nA, (b) $U_{sample}$ = +0.05 V / $I_t$ = 0.15 nA, (c) $U_{sample}$ = +0.1 V / $I_t$ = 0.15 nA (a) before and (b) after dosing 0.0015 L CO and (c) 0.003 L CO. The CO interacts first with defects (black and white squares, marked by blue and purple circles, respectively). Once all defect sites are saturated, the CO adsorbs at the clean surface (red circle).

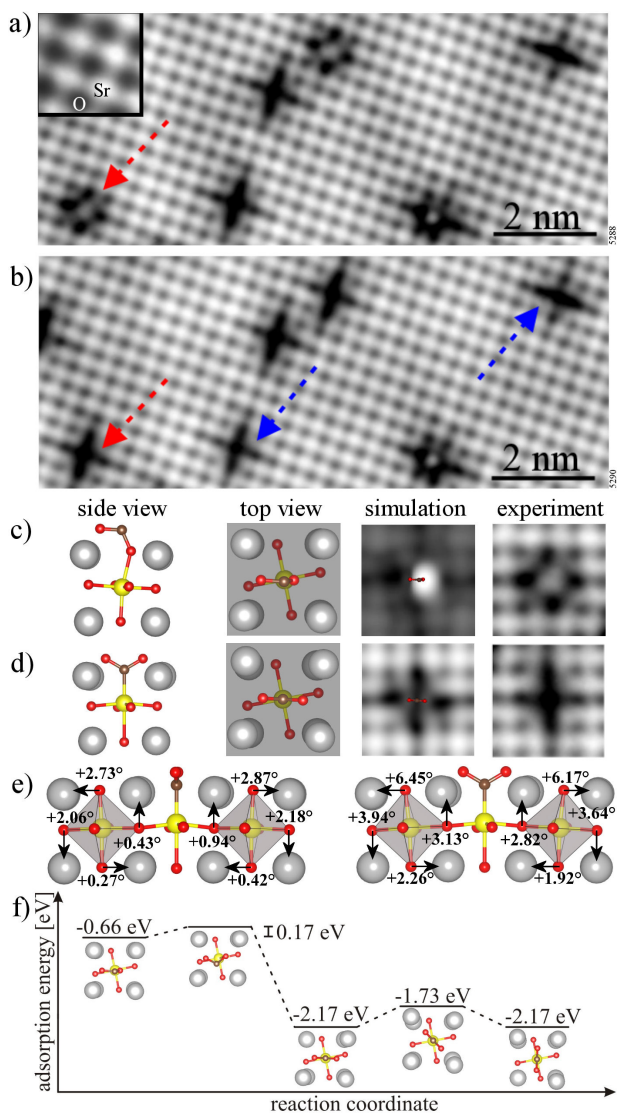

**Fig 3. Configurations of adsorbed CO**. (a, b) STM images: 11×4.5 nm², $U_{sample}$ = +0.2 V, $I_t$ = 0.15 nA, $T$ = 78 K. The insert shows a calculated STM image of the clean surface; Sr atoms are imaged bright. The red arrows mark a CO molecule that is transformed from a physisorbed precursor state into a metal carboxylate (Ru-COO) species, imaged as a large cross. The blue arrows mark two crosses, rotated by 90°. (c, d) Structure model, simulated, and experimental STM images of (c) the precursor and (d) the carboxylate. In both cases Sr and Ru atoms are pushed away by the adsorbed CO. Note that two equivalent orientations of the OCO group are possible, resulting in the 'rotated' crosses. (e) Local lattice distortion caused by the carboxylate (f) Binding energies for the various adsorption configurations.

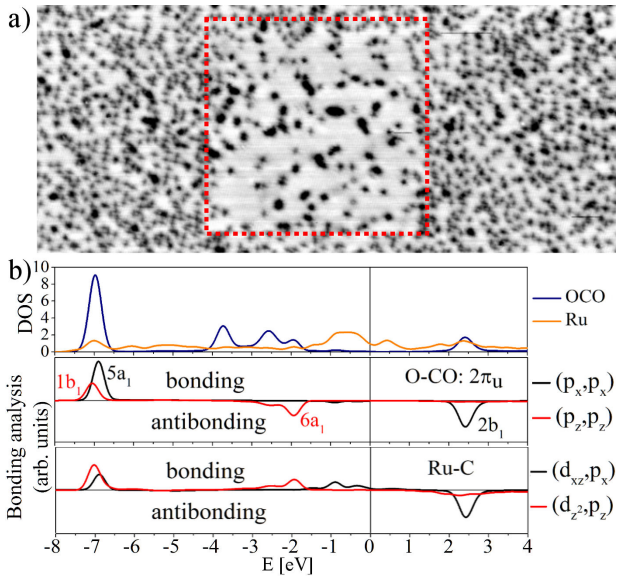

**Fig 4.** (a) STM image: 65×34 nm², $U_{sample}$ = +0.2 V, $I_t$ = 0.15 nA, $T$ = 78 K. Scanning at +2.7 V removes the CO from the scanned area (red, dotted frame). The black spots inside the square were already present before removing the CO and are attributed to surface defects. (b) DOS of chemisorbed OCO. Note the LUMO at +2.4 eV, tunneling into these states weakens the O-CO as well as the Ru-C bond.